\documentclass[12pt,preprint]{aastex}
\usepackage{amssymb}
\usepackage{txfonts}

\begin{document}

\shorttitle{Spectropolarimetry of Fraunhofer lines} \shortauthors{Qu
et al.}

\title{Spectropolarimetry of Fraunhofer lines in local upper solar atmosphere}

\author{Z.Q. Qu$^{1,2}$, L. Chang$^{1,2}$, G.T. Dun$^{1}$, X.M. Cheng$^{1}$, C. Fang$^{3}$, Z.Xu$^{1}$, D.Yuan$^{4}$, L.H. Deng$^{1}$,
X.Y. Zhang$^{1}$\\
1. Yunnan Astronomical Observatories, CAS, Tinwentai road, Kunming, Yunnan 650216, China\\
2. School of Astronomy and Space Science, University of Chinese
Academy of Sciences, Yuquan road, Chaoyang district, Beijing,
China\\
3. School of Astronomy and Space Science, Nanjing University, Hankou
Road, Nanjing, Jiangsu, China\\
4. Institute of Space Science and Applied Technology, Harbin
Institute of Technology, Shenzhen, Guangdong, China}

\begin{abstract}
Spectropolarimetric results of Fraunhofer lines between 516.3nm and
532.6nm are presented in local upper solar chromosphere, transition
zone and inner corona below a height of about 0.04 solar radius
above the solar limb. The data were acquired on Nov.3, 2013 during a
total solar eclipse in Gabon by the prototype Fiber Arrayed Solar
Optical Telescope(FASOT). It is found that the polarization
amplitudes of the Fraunhofer lines in these layers depend strongly
on specific spectral lines. Fraunhofer line at MgI$b_{1}$518.4nm can
have a polarization amplitude up to 0.36$\%$ with respective to the
continuum polarization level, while the polarizations of some lines
like FeI/CrI524.7nm and FeI525.0nm are often under the detection
limit 6.0$\times 10^{-4}$. The polarizations of the Fraunhofer
lines, like the emission lines and the continuum, increase with
height as a whole trend. The fractional linear polarization
amplitudes of inner F-corona can be close to those of inner
E-corona, and in general larger than those of inner K-corona.
Rotation of the polarization direction of Fraunhofer line is often
accompanied with variations in their polarization amplitudes and
profile shapes. It is also judged from these polarimetric
properties, along with evidences, that neutral atoms exist in these
atmospheric layers. Thus the inner F-corona described here is
induced by the neutral atoms, and the entropy of the inner corona
evaluated becomes larger than those in the underneath layers due to
more microstates found.
\end{abstract}


As a common knowledge, solar atmosphere is stratified and formed by
the photosphere, chromosphere, transition zone and corona as the
heliocentric height increases. In this paper, our study is related
to its upper atmosphere containing upper chromosphere, transition
zone, as well as inner corona. In these layers, the temperatures of
most particles continue to increase with height till the middle
layers of the inner corona. This leads to a conclusion hitherto that
all the particles coming from the below layers are heated and
ionized in the corona(Golub and Pasachoff,1997; Aschwanden,2015).
However, such a view will be changed as demonstrated in this paper.

In order to diagnose the physical conditions, visible light
irradiated from quiet solar corona is divided into three parts
according to their forms in the literature till now(Golub and
Pasachoff,1997; Aschwanden,2015). They are respectively continuum
part called K-corona mainly yielded by free electrons scattering the
photospheric radiation, E-corona shaped by distribution of emission
lines attributed to bound-bound transitions in particles, and
F-corona, initially discovered by Moore(1934) and Grotrian(1934),
formed via scattering or/and diffracting radiation of
absorption(Fraunhofer) lines irradiated from solar photosphere by
the dust grains distributed mainly in the ellipsoid around the
sun(Lamy et al.,2022; Burtovoi et al.,2022), as an extended corona.
Such a kind of scattering is called Mie scattering(Lietzow,2023).
After efforts of more than forty years, Koutchmy and his
collaborators(Koutchmy et al., 1973; Koutchmy et al., 2019)
demonstrated that the projective height(elongation) of the
Fraunhofer lines can be very low above the solar limb from eclipse
observations. On the other hand, by photometry, its connection to
zodiacal light was established (Lamy et al., 1992; Boe et al., 2021;
Lamy et al., 2022; Burtovoi et al., 2022).

Polarimetry especially Stokes spectropolarimetry can provide the
irreplaceable tool to diagnose the physical conditions. In fact,
great efforts for polarimetry of the corona especially F-corona and
their theoretical predictions as well as interpretations form a
history longer than a century, and solar eclipses have been favorite
since they provide the clean environments of minimum scattering and
stray light via telluric atmosphere, or observations are performed
in space out of the earth with the minimum light pollution. These
polarimetries were dominantly concerned of large scales beyond a
height of three solar radii(3$R_{sun}$) above the solar limb around
the sun from broad band observations based on filters. Therefore,
polarizations of the F-corona in most cases were not directly
obtained but indirectly derived from measured total polarization
$P_{P+K}$ and calculated $P_{K}$ induced from distribution of the
free electrons, with knowledge of total intensity and K-coronal
intensity(Hulst,1950). One of main scientific goals of these
polarimetries is to separate the K- and F-components of the corona
and then calculate the electron density as a function of distance
from the sun, meanwhile the E-corona is often ignored, though it can
be present in the broad band observations.

Therefore, the direct polarimetry of the F-corona was very rare. To
our knowledge, Ohman(1947) carried out a spectropolarimetry of solar
corona with a lowest height to 0.3$R_{sun}$(solar radius) above the
limb during 1945 July 9 total solar eclipse. Due to the low spectral
resolution and poor signal-to-noise ratio, it was hard to recognize
even the Fraunhofer lines from the continuum. This led to that his
measured F-corona polarization $P_{F}$ was very uncertain according
to their analysis. It gave an extrapolation result of 0.42 at height
of 0.13$R_{sun}$ and 0.11 at 0.19$R_{sun}$ in G-band listed in his
Table XIII, and concluded that the polarization of the Fraunhofer
lines was smaller than that of the continuum. Later, Blackwell and
Petford(1966) presented their polarimetric results obtained during
1963 July 20 solar eclipse. In their Table I, percent polarizations
of the F-corona were derived from 0.19$\%$ at height of
9$R_{sun}$(solar radius) to 0.65$\%$ at height of 15$R_{sun}$ when
half-width of the filter bandpass of the polarization channel was
set to be 11.6nm. And they constructed an outer solar corona model
within which the F-corona polarizations ranged from a very small
value 0.05$\%$ at height of 5$R_{sun}$ to 2.84$\%$ at height of
40$R_{sun}$, while the polarizations of K-corona were modeled from
59.7$\%$ to 61.4$\%$. In his paper, Mann(1992) concluded that dust
particles close to the sun influence the coronal brightness
strongly, and the calculated F-corona polarization ranges from
0.09$\%$ at $8R_{sun}$ to 0.84$\%$ at 16$R_{sun}$, with zero level
at $5R_{sun}$ as the boundary condition. Polarization variation with
distance from the sun is fitted by a function of $(r/r_{0})^{-p}$
($p=2.65-2.8$) valid till 0.2AU, and $r_{0}$ is the distance between
the sun and the earth, i.e., 1AU. They demonstrated that a small
polarization by circum-solar particles is necessary to fit their
brightness observations. From near-sun observations of the F-corona,
Howard and his cooperators(2019) found the intensity decease of the
F-corona at short elongations without any polarization data, which
was thought to be suggestive of the long-sought dust free zone.
Based on data acquired during 2019 July 2 total solar eclipse, Boe
and his cooperators(2021) presented the color and brightness of the
F-corona, and they found that the total brightness of the F-corona
higher than theoretically expected in the lower corona, and deduced
that the F-corona is slightly polarized. One outstanding feature of
the F-corona is pointed out that the relative line intensities among
Fraunhofer lines are very close to their photospheric
counterparts(Koutchmy et al., 1973; Koutchmy et al., 2019).
Recently, Burtovoi et al.(2022) derived the maps of the F-corona via
time correlation of total and polarized visible light images
obtained from SOHO/LASCO-C2 (Brueckner,1995; Domingo,1995). The
derivation was based on a new analysis of the evolution of the total
and polarized brightness images. From the intensity contours of
their Fig.15, approximately elliptic shapes of the derived F-corona
were found.

In fact, the theories were more precedent to the polarimetry. Early
in 1879, Schuster(1879) tried to theorize solar coronal polarization
due to Rayleigh scattering. He got a conclusion that the linear
polarization should increase with distance from solar center if
distribution of the scattering particles obeys inverse power of the
distance from the sun. In his Tables, the polarization values were
listed respectively, derived from different inverse power
distribution, and correspondingly a great range of polarization
values were obtained. It is noticeable that Schuster did not take
into account the limb-darkening. Half century later, Minnaert(1930)
studied the polarization of coronal continuum (i.e., K-corona) in
1930 due to scattering of photospheric light by free electrons,
assuming that the electron density distribution could be described
as a power function $r^{n}$ of radius $r$, and he found that the
polarization depends slightly on the wavelength, and polarization
amplitude at least 16.4$\%$ shown in his Table 6 was expected at
solar limb from assuming an electron density variation with the
distance of $r^{-8}$ from the sun. However, he found that the
measured values were much smaller than these predicted, and assumed
that the corona does not only scatter the photospheric light but
also there were some other particles emanating radiation of the
continuum. More than twenty-five years later, Grotrian(1956) gave
the interpretation of the Rayleigh scattering of solar photospheric
light by the particles, and tens of percent polarization amplitudes
were estimated. The most memorized theory about the polarization in
the early stage was given by van de Hulst(1950). He formulated the
polarization of the corona, but found his theoretical results of
K-corona greater than observed values then, and thus ignorance of
F-corona polarization might lead to discrepancies. In his theory,
the polarization was wavelength-independent in a narrow band if only
Rayleigh scattering was considered. A curve describing variation of
the F-corona percentage polarization with elongation was given by
Blackwell(1956) from his theoretical calculation. He showed an
ignorable value close to solar limb but high percentage polarization
of about 15$\%$ at elongation $60^{\circ}$, and his actual
measurement from an aircraft during a total solar eclipse led to
much higher values. For instance, the theoretical prediction was
less than one percent at elongation of $5^{\circ}$ but his
polarimetry resulted in a value of 2.8$\%$, which was used for
identification of the zodiacal light as the outer F-corona. Very
recently, since the brightness of F-corona in the low corona is
measured to be more intense than expected by assuming unpolarized
F-corona, Boe and his cooperators(2021) are doubting the assumption.
A method for separating K-corona and F-corona is proposed by
Burtovoi et al.(2022) via acceptance of less than 0.06$\%$ linear
polarization amplitudes of the F-corona below 6$R_{Sun}$, calculated
by Blackwell and Petford(1966). Nevertheless, these theoretical
predictions differed considerably due to their different assumptions
and many inconsistencies among the polarimetric results and the
theoretical calculations lead to chaos in this literature.

\section{Observation and Data reduction}

 The spectropolarimetric data were obtained during 2013
total solar eclipse observation by the prototype Fiber Arrayed Solar
Optical Telescope(FASOT)(Qu, 2011; Qu et al.,2014), observed at
Bifoun of Gabon on Nov.3, 2013. The observation aimed at real time
spectroimaging polarimetry within a band from 516.3nm to 532.6nm
containing multiple spectral lines(Qu et al.,2017; Qu et al.,2022),
and data used here are only a small part of the dataset. The
instrument and some observational details were introduced in the
last two cited papers, but necessary details are described again
below in order to make a brief and clear picture about the spectral
images shown respectively in top panels of following Figures 2-6. In
our last two papers, we presented respectively spectro-imaging
polarimetry of the green coronal line(Qu et al.,2017)(referred to
Paper I hereafter), and spectropolarimetry of all the emission lines
observed(Qu et al,2022)(referred to Paper II hereafter), but
remained polarimetric demodulation and analysis of the Fraunhofer
lines and continuum observed in the upper solar atmosphere in this
paper due to its abundant content.

The approximate positions of observational field of views(FOVs) of
sample data documented with SE196-2, SE196-4, SE204-3 and SE205-1
are indicated by the white squares (not to scale) artificially
superposed on corresponding eclipse monitoring images respectively
shown in three panels of Figure 1, where the bright sickles were
white-light images of the sun blocked by the moon in process of
approaching the second contact. Each of the two FASOT integral field
unit(IFU) head structures defining the same field of view(FOV) is
formed by 5$\times$5 microlens array receiving radiation from
spatially resolved points in the sun, coupled with fiber array
behind one by one and coded in the bottom right panel of Fig.1. Each
spatial volume covers 2 $\times$ 2 arcseconds or approximately
1500km$\times$1500km cross-section transversally in the sky plane.
The fibers forming the array in the IFU head are re-weaved into one
pseudo slit dispersed by a spectrograph. The spectral images on top
panels of Figures 2-6 are obtained after dispersion of the fifty
beams coming out of the fifty fibers contained in the slit. It is
worthy to note that in order to carry out the accurate linear
polarimetry, the light beam originated from each of the solar
spatial volume is split by the polarizing beam splitter into two
beams with opposite polarization states denoted by $'a'$ and $'b'$
in spectral images on top panels of Figs.2-6, received by the pair
of IFU. Thus there are 50 rather than 25 spectra as relatively
bright horizontal strips in each top panel of these figures. The
intensity and fractional linear polarization are demodulated from
each couple of fibers respectively. The two spectra of each couple
are arranged symmetrically about the central dark horizontal belt
between the two twenty-fifth couple spectra. For instance, the two
beams of the first couple($1a$ and $1b$) are placed respectively on
the bottom and the top rows, and the other couples move to the
cental line orderly, as illustrated in the right columns of the top
panels of these figures. The intensities of $'a'$ and $'b'$ of one
couple can be described by $1/2(I+Q)$ and $1/2(I-Q)$ respectively
multiplied by the extinction factors caused by the time-dependent
telluric atmospheric seeing and different light paths within the
telescope(see Papers I and II). A geographic direction indicator is
also plotted on the left side in the bottom right panel of Fig.1
together, and the positive linear polarization Stokes $Q$ is defined
along the North-South orientation.

\begin{figure}
\flushleft
\includegraphics[width=16cm,height=9.cm]{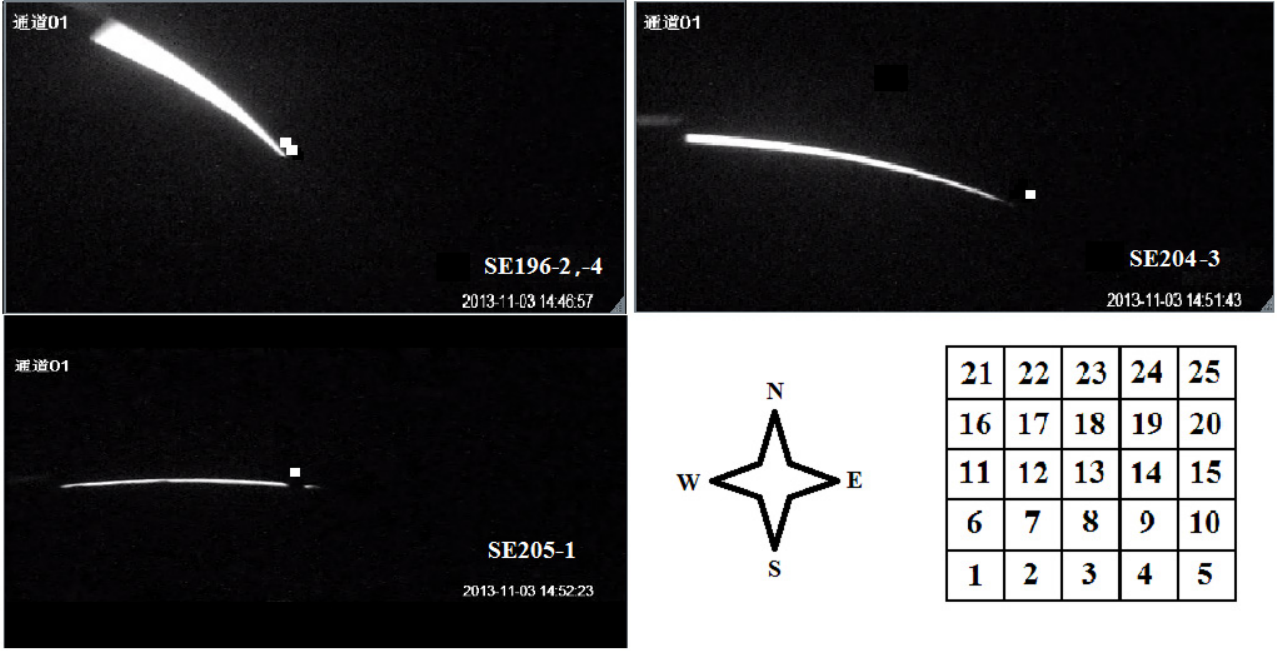}
\caption{\footnotesize Approximate locations of fields of view
(FOVs) of the polarimetric data samples analyzed in the text,
directional indicator and alignment configuration of the
lenslet-coupled fiber array as the head of Integral Field Unit (IFU)
of the prototype FASOT. Top two and bottom-left panels: bright
sickles were solar crescents formed by lunar occultation in process
of the eclipse. Artificial white squares corresponding to the
bottom-right square icon superposed in these monitoring images, not
to scale, indicate approximately estimated positions of FOVs. The
monitoring images of SE196-2 and SE196-4 are combined on the top
left panel, that of SE204-3 is plotted in the top right panel, and
that of SE205-1 in the bottom left panel. The location of SE196-2 is
a little higher than SE196-4. Bottom right panel: the directional
indicator placed on the left, and on the right is the alignment
configuration of 5$\times$5 spatial points of the FOVs indicated as
the white squares in the other panels of this figure. } \label{}
\end{figure}

The polarimetric demodulation technique was described in Paper I and
then improved in Paper II, and again we adopted the method described
by Eq.(3) in Paper II. However, unlike data reduction of emission
lines in Paper II, where most of the polarimetric demodulations were
executed for each resolved spatial point with an intensity threshold
above the continuum level. The demodulation of the Fraunhofer lines
in the upper solar atmosphere will not become feasible especially at
the line centers with the least residual intensities for each
resolved spatial point, due to the noise fluctuation typically of
tens of readout photon counts leading to polarimetric noise level of
6.0$\times 10^{-3}$, a value greater than all these demodulation
signals in the following. Therefore a spatial binning over rows of
the FOV(referred to the diagram in the bottom right panel of Fig.1)
becomes necessary to obtain the reliable polarimetric results
without demodulation threshold. This leads to reduction of the
spatial resolution but enhancement of the polarimetric
signal-to-noise ratio, and reliable polarization information of the
continuum spectra becomes also available simultaneously.

The Stokes polarimetric result of the quiet sun region at the solar
disk center before the eclipse can be used to provide a standard
photospheric Fraunhofer lines. Otherwise, in order to detect the
effect of above-mentioned binning and check the accuracy of the
polarimetry, we demodulate the photospheric data named SE90-1
acquired at the disk center in the quiet sun region with a small
fractional aperture of the telescope used before the eclipse. Three
sets of demodulated intensity $I$ and percent linear polarization
represented by Stokes $Q/I$ profiles are depicted in Fig.2 for
specifying the binning effect. The intensity is in unit of the
readout photon counts(cts) of detector throughout the paper.

\begin{figure}
\flushright
\includegraphics[width=14.1cm,height=3.2cm]{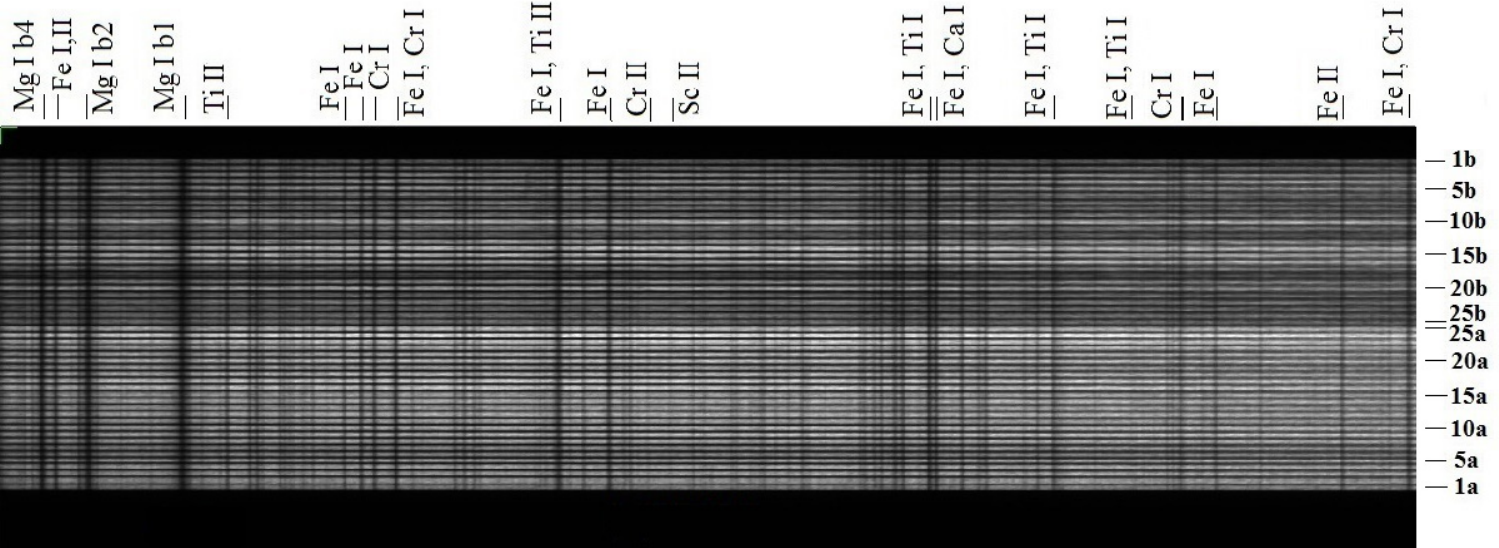}\\
\flushleft
\includegraphics[width=16cm,height=5.0cm]{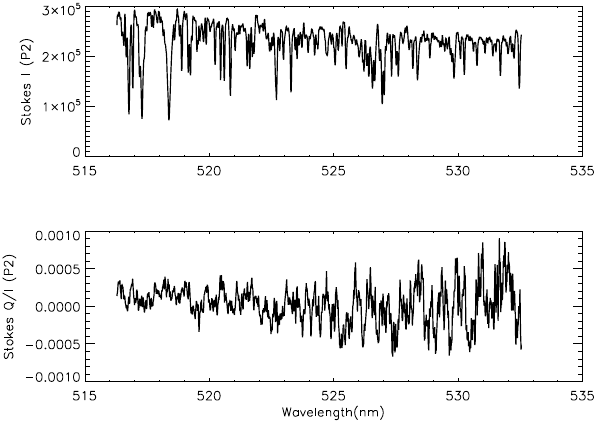}\\
\flushleft
\includegraphics[width=16cm,height=5.0cm]{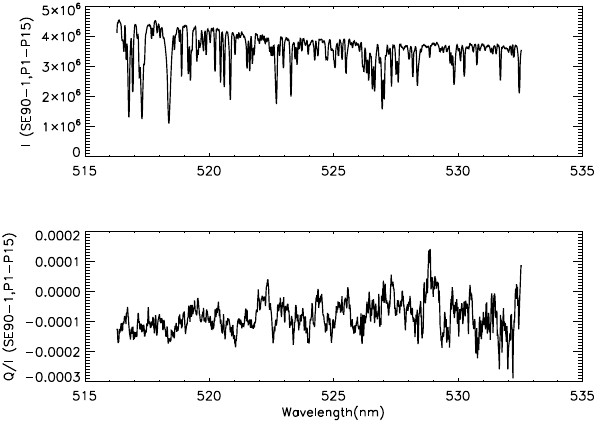}\\
\flushleft
\includegraphics[width=16cm,height=5.0cm]{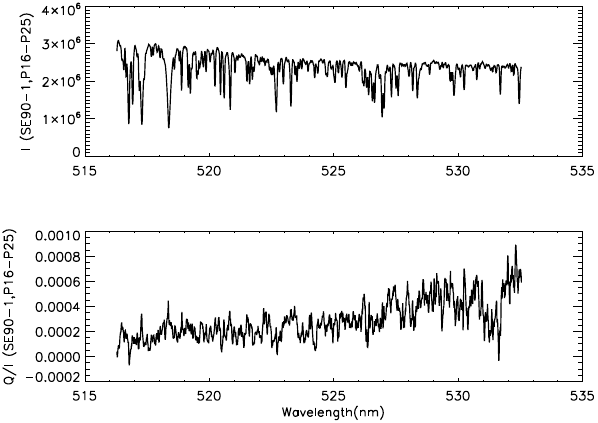}\\
\caption{\footnotesize Sample polarimetric results acquired at the
disk center before the eclipse. The demodulated intensity and the
fractional linear polarization $Q/I$ profiles without and with
binning performances are plotted below the raw spectral image to
show the enhancement of the polarimetric accuracy with the spatial
binning. The symbols above the top image indicate the line producers
with their ionization state, and the numbers attached with letters
'a' and 'b' in the rightmost column reflect ordinary and
extraordinary light beams originated from the incident volume array
described in the bottom-right panel of Fig1. The intensity $I$ is
obtained via adding the dual beams with the same numbers, and $Q/I$
is demodulated also from the dual beams via a technique called
Reduced Polarimetric Optical Switching(RPOS), and hereafter in the
figures. }
\end{figure}

Top panel of Fig.2 provides the raw spectral image of the
photospheric Fraunhofer lines within the observational band from
516.3nm to 532.6nm. According to the spatial symmetry, the
polarization here should be zero. Since we have enough photon budget
in this photospheric case, a sample demodulation of the second
resolved spatial point P2 without binning is displayed in the second
and third panels(counted from top to bottom and hereafter). The
fractional polarization noise does not exceed 1.0$\times 10^{-3}$
with respective to(wrt) the continuum level. Then we perform the
binning over the three lower rows of the FOV, i.e., from point one
to point fifteen(indicated by P1-P15 and similar in the following),
shown in the fourth and fifth panels. It is markable that the
polarimetric noise is reduced to be below 3.0$\times 10^{-4}$ wrt
the continuum polarization level. Finally, we present the result
after binning P16-P25. The polarimetric noise level is smaller than
the single point P2 but greater than the binning P1-P15 which
contains more spatial points. However, since the scattering and
stray light was greatly depressed, the polarimetric accuracies will
be increased correspondingly for these data acquired close to the
totality. It is noteworthy that the polarization of the quiet sun
region at the disk center is used for polarization zero
calibration(Bianda,1998) as pointed out in Paper II.

\section{Spectropolarimetric results of Fraunhofer lines during the
eclipse}

In this section, we present four polarimetric cases respectively in
an order according to their acquisition times. Two samples show
Fraunhofer lines coexisting with emission ones in the same FOVs but
without any coronal line, and the other two are cases of Fraunhofer
lines present with the emission lines including the green coronal
line.

As pointed out previously, the residual intensity of a Fraunhofer
line of single spatially resolved point is too faint for us to get
enough polarimetric signal-to-noise ratio, we have to carry out
divisions of the FOV via the spatial binning to promote the ratio.
The division is classified into two kinds in this paper. One is
executed in such a way that the binning is performed over
respectively the lower three rows and then the upper two rows of the
FOV. The other is accomplished respectively over the lower two rows
and then the upper three rows according to the need(cf. bottom-right
panel of Fig.1). Otherwise, binning over all these spatial points in
the FOV is performed so that we can see further polarimetric result
over a larger spatial volume and furthermore how spatial integration
influences the polarimetric consequences. Finally, it is valuable to
bear in mind that intensity of one spatial point is an integration
along line-of-sight of radiative sources with different heliocentric
heights, and polarization amplitudes of spectral lines described
below are generally referred to be with respect to(wrt) their
adjacent continuum polarization level.

\subsection{Spectropolarimetric results of Fraunhofer lines without the coronal line}

The Fraunhofer lines are dominantly originated from the photosphere
in appearance of absorption under the background of the thermal pool
formed by the frequent particle collisions(Mihalas,1969), and the
absorption dominates the emission. However, they can be also
produced in local upper regions where the collision plays still an
important role or observed via Rayleigh scattering. Whether these
Fraunhofer lines can survive in the inner corona is just what we
want to investigate. In this subsection, we select two cases named
SE196-2 and SE196-4 within one set of data SE196, which contains six
frames acquired with an exposure time of three seconds for each
frame, about four and half minutes before the totality, when the
bright solar crescent formed by lunar occultation was still
prominent as seen from the top left panel of Fig.1, where the FOV of
SE196-4 locates a little lower than SE196-2 above the local solar
limb. In fact, there are more than twenty frames acquired with
abundant spatial distribution patterns of the Fraunhofer and
emission lines together without the coronal line present in the same
FOVs. The distribution patterns of the Fraunhofer lines in the FOVs
of these two selected samples can be representatives. They are
regarded to be in upper chromosphere and extended into the
transition zone judged from their presence above their chromospheric
emission counterparts such as neutral magnesium triplet( Mg{\sc I}
$b_{1}: 518.4nm, b_{2}: 517.3nm$ and $b_{4}:516.7nm$), and the
transition zone emission lines represented by once ionized iron
emission lines, such as FeII531.7nm. It is noteworthy that no
regularly geometrical boundaries between the corona and transition
zone, or even between the transition zone and chromosphere,  as a
common sense in the literature. Therefore the emission lines can
extend to different heights in different regions, the identification
of the forming locations of the Fraunhofer lines concerned depends
on their distribution compared with these emission lines.

\subsubsection{SE196-2}

The first case named SE196-2 is depicted in Fig.3. The raw spectral
image is shown in the top panel. It is seen that most Fraunhofer
line appearances can be seen in these points above those yielding
their emission line counterparts, but the transformation from line
emission to depression depends on spectral lines at different
spatial points. This can be ascribed to their different emission,
absorption and scattering along the line-of-sight, and differs from
the scenario caused purely by dust scattering. Especially, there are
few weak Fraunhofer lines keeping their absorption appearance in
almost the whole FOV, such as CaI526.0nm, FeI521.6nm,
CrI/FeI/FeII527.3nm, CrI529.8nm and CrI530.1nm.

The second and third panels(counted from the top to the bottom and
hereafter) present respectively the intensity $I$ and corresponding
fractional linear polarization $Q/I$ profiles of a spatial domain
covering the three lower rows of the FOV(referred to bottom right
panel of Fig.1). The Fraunhofer lines are hard to be seen in the
second panel, due to depression by the emission lines via linear
intensity contrast. But this is not the case for their $Q/I$
profiles. Besides the prominent polarizations at most of the strong
emission lines, the very weak Fraunhofer lines of mixed
CrI/FeI/FeII527.3nm and CrI529.8nm are polarized with $Q/I$
amplitudes comparable to those of the emission lines with largest
$Q/I$ polarizations. Some of other Fraunhofer lines like these at
521.3nm, 523.2nm, 527.9nm and 531.5nm gain distinguishable $Q/I$
amplitudes comparable to these of some medium strong emission lines.
But polarizations of most of the Fraunhofer lines merge still in
noise.

When the spatial binning is performed over the two higher
layers(P16-P25), shown in the fourth and fifth panels, the
Fraunhofer lines become dominant now, and their relative line depths
are changed from the photospheric counterparts shown in Fig.2.
Polarimetrically, FeI/CrI532.4nm line attains the greatest amplitude
of about 0.32$\%$ with respective to(wrt) the local continuum level
0.28$\%$. The magnesium triplet and these lines around respectively
526.6nm, 527.0nm, 529.0nm, while lines at 529.8nm and 531.8nm obtain
considerable polarizations, but many of the Fraunhofer lines
including some strong ones remain to own no detectable polarization
signals.

\begin{figure}
\flushright
\includegraphics[width=14.2cm,height=3.2cm]{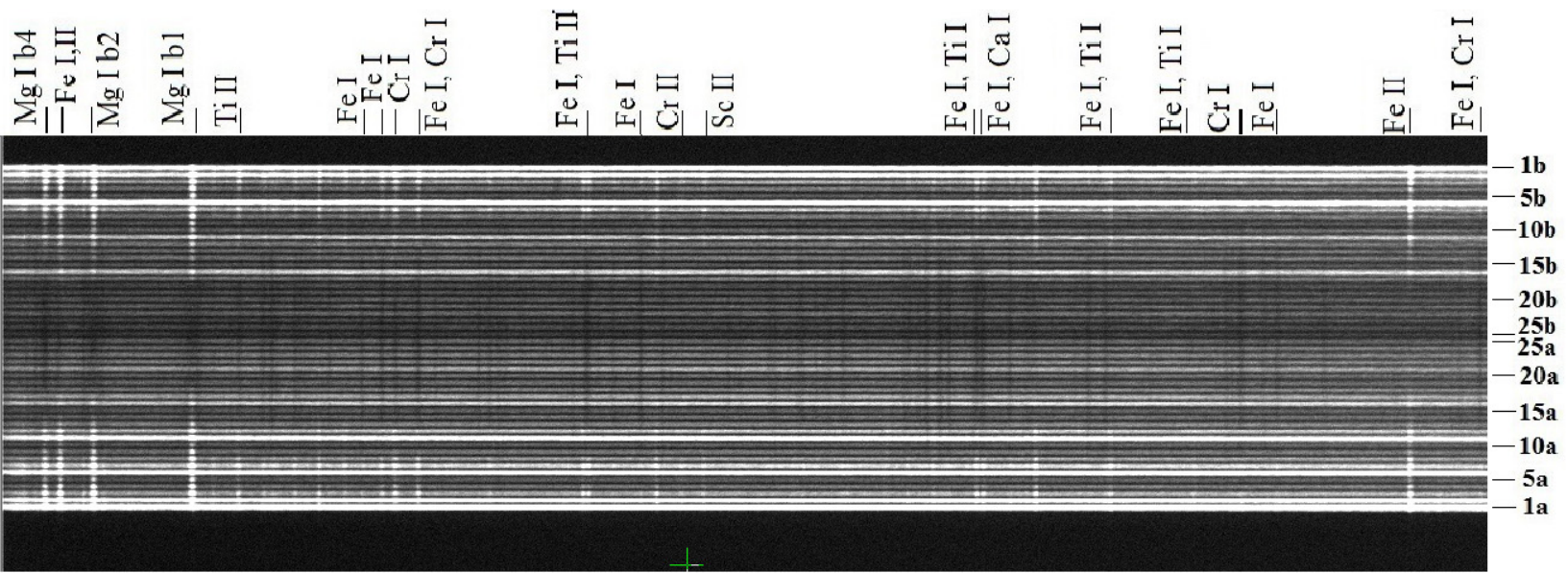}\\
\hspace{3.8cm} \flushleft
\includegraphics[width=16cm,height=5.2cm]{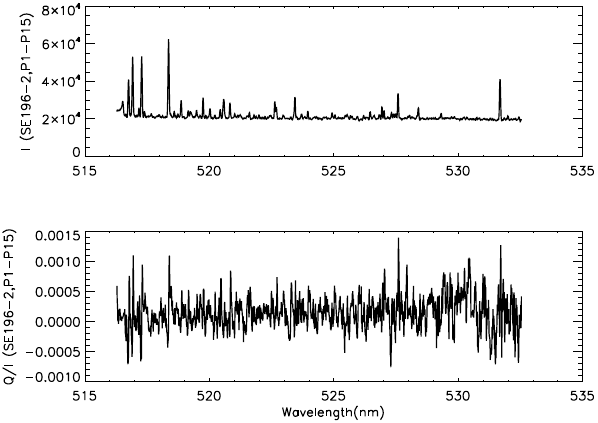}\\
\hspace{3.8cm} \flushleft
\includegraphics[width=16cm,height=5.2cm]{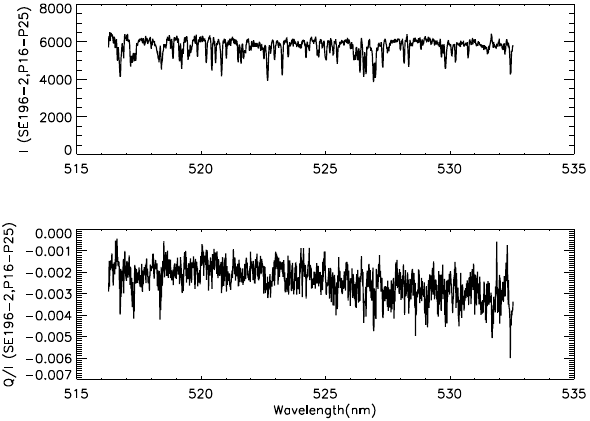}\\
\hspace{3.8cm} \flushleft
\includegraphics[width=16cm,height=5.2cm]{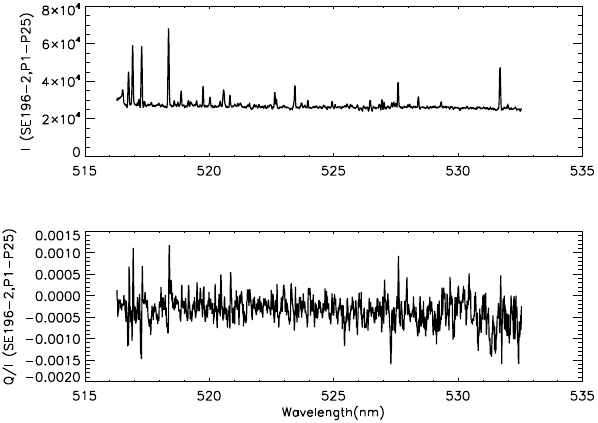}
\caption{\footnotesize SE196-2. Top panel: the raw spectral image.
Lower panels: the intensity $I$ and fractional linear polarization
$Q/I$ profiles obtained from polarization demodulation after spatial
binning over P1-P15(cf. bottom panel of Fig.1), P16-P25 and the
whole(P1-P25) of the FOV respectively. }
\end{figure}

After the whole FOV binning is performed as shown in the two bottom
panels, it is evident that both the $I$ and $Q/I$ profiles look much
closer to those in the lower layers. This is understandable since
that much greater number of photons and polarized photons are
contributed there. However, the contribution of these polarizations
below the continuum level from the higher layers are observable. For
instance, the large negative amplitude of the 532.4nm line is more
ascribed to the upper layers. And the most prominent positive $Q/I$
value locates at 527.6nm in the lower layers but now transfers to
MgI $b_{1}$ line. It becomes evident that the polarimetric noise
level decreases, as more easily witnessed in the subband with
wavelengths larger than 532.0nm.

Therefore, it becomes evident that the polarizations of the
Fraunhofer lines, like their emission counterparts also described in
Paper II, depend on specific lines. On the other hand, as a whole
trend, the polarization amplitudes of these lines increase when they
are transformed from the emission lines to the Fraunhofer ones with
detectable polarization. In fact, the polarization plane rotation
can be seen in not all but many lines that in the lower layers there
are positive polarizations in the red wings and negative ones in the
blue wings but in the higher domain, the polarization becomes
dominantly negative.

\subsubsection{SE196-4}

The second case is SE196-4 with its FOV lower than SE196-2. It shows
evidently another distribution pattern of alternative domination of
the emission and absorption in different lines from Fig.4. Most
lines own emission appearance in the lowest rows of the FOV but some
lines turn to be Fraunhofer lines in higher layers like
FeI/NiI519.2nm, FeI520.2nm, FeI/CrI520.4nm, CrI520.6nm and
CrI/FeI520.8nm, while more lines especially the strong ones keep a
form of emission in the whole FOV, such as the quartet(chromospheric
magnesium triplet and the mixed FeI/FeII516.9nm, and hereafter),
transition zone lines of FeII531.7nm. However, most of these
emission lines are not of pure emission but with absorption feature
in their far wings in detail. On the other hand, some weak
Fraunhofer lines like FeI526.3nm, CaI526.6nm and CrI/FeI532.4nm
maintain their absorption form in the whole FOV, which implies their
photospheric origin. It is seen that the quartet is the strongest,
then followed by the two mixed emission FeII531.7nm lines. The
relative intensities among these Fraunhofer lines in the higher
layers(cf. the fourth panel of Fig.4) are not consistent with those
of the photosphere. For instance, CaI/FeI527.0nm becomes much weaker
than those in Fig.2 compared with its surrounding lines.

The second and third panels of Fig.4 show respectively the intensity
$I$ and fractional linear polarization $Q/I$ of a spatial domain
covering the three lower rows of the FOV(referred to bottom right
panel of Fig.1), i.e., from the first volume to the fifteenth
one(P1-P15). It is very hard to see the Fraunhofer lines, because
that they are greatly depressed by the emission lines via contrast.
However, it can be seen that the very weak Fraunhofer lines of mixed
CrI/FeI/FeII527.3nm and FeI/CrI/NdII529.4nm are polarized with
prominent $Q/I$ amplitudes. And groups of lines around 517.6nm,
521.4nm, 523.2nm, 525.5nm, 529.2nm and 531.4nm containing very weak
Fraunhofer lines yield evident $Q/I$ valleys. It is valuable to note
FeI/CrI532.4nm line with a very weak emission appearance, its
polarization is prominent. Its polarization turns to be stronger in
the higher elongations where it turns to be a Fraunhofer line. High
signal-to-noise ratios of polarizations of the quartet and the
transition zone emission lines can be seen. Some of them have $Q/I$
amplitudes close to 0.10$\%$. These $Q/I$ profiles of the quartet
and other lines are approximately antisymmetric about their line
centers. In detail, the polarizations in the blue wings of these
lines are negative(along East-West direction) and those in the red
wings positive(along North-South direction). It is easily deduced
that if one carries integration along dispersion over the spectral
lines individually, the integrated polarization amplitudes will
become much weaker or even canceled out. This specifies the
importance of spectral resolution in polarimetry. It is also noted
that this kind of $Q/I$ profile is not shared by some other lines
such as FeII531.7nm that has an outstanding single peak in its $Q/I$
profile. Finally, some Fraunhofer lines have no detectable
polarization like 522.3nm, 524.2nm, 524.8nm, 526.3nm and 526.6nm.
Therefore, the $Q/I$ profile configuration depends on spectral lines
and space.

\begin{figure}
\flushright
\includegraphics[width=14.0cm,height=3.2cm]{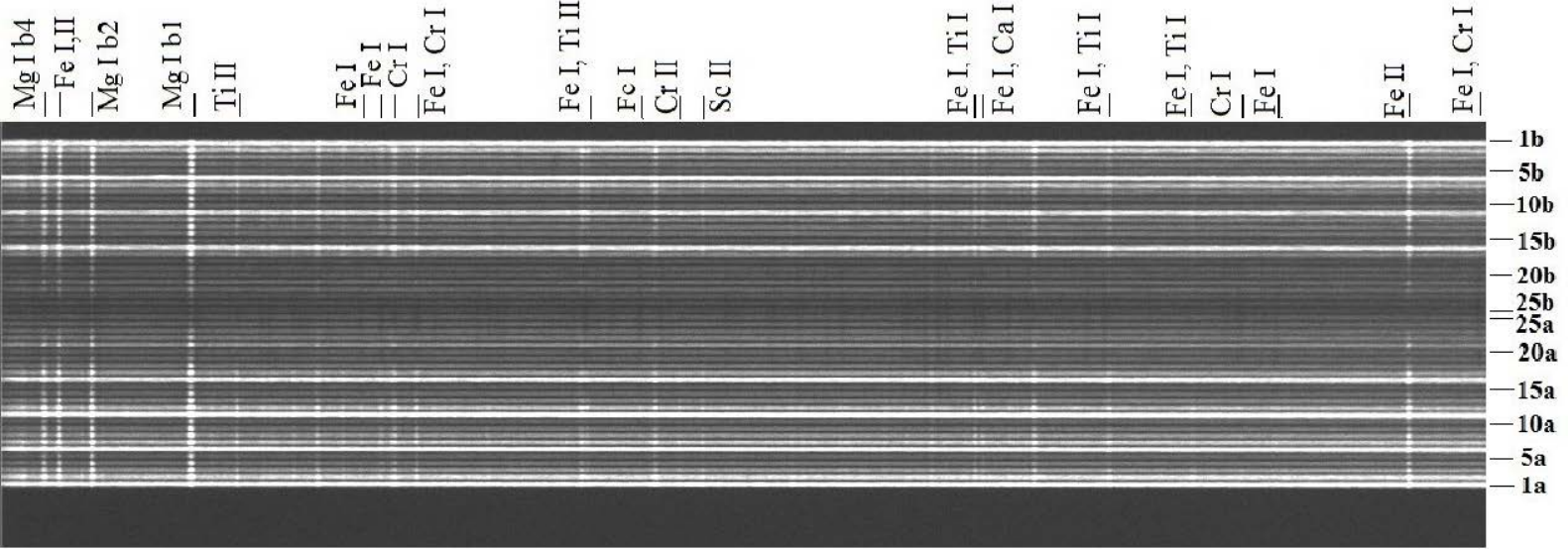}\\
\hspace{3.8cm} \flushleft
\includegraphics[width=16cm,height=5.2cm]{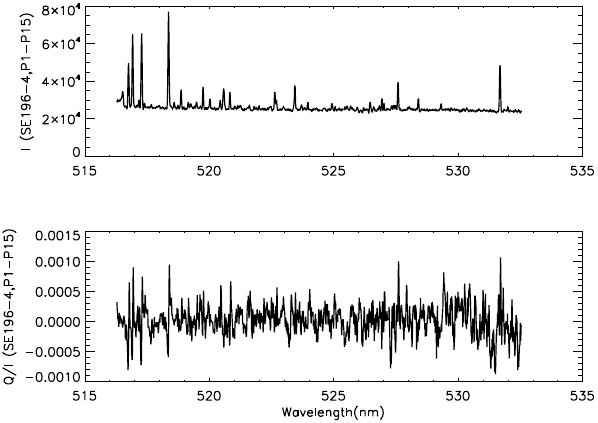}\\
\hspace{3.8cm} \flushleft
\includegraphics[width=16cm,height=5.2cm]{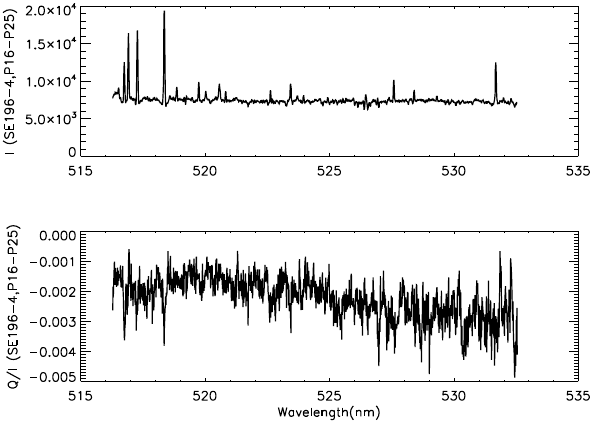}\\
\hspace{3.8cm} \flushleft
\includegraphics[width=16cm,height=5.2cm]{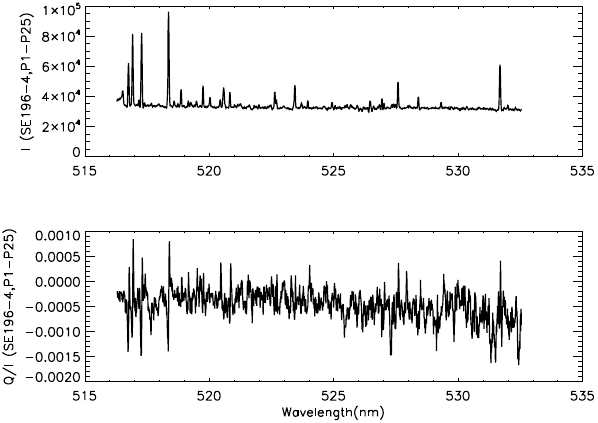}
\caption{\footnotesize SE196-4. Top panel: the raw spectral image.
Lower panels: the intensity $I$ and fractional linear polarization
$Q/I$ profiles obtained from polarization demodulation after spatial
binning over point one to point fifteen(P1-P15), P16-P25 and all the
points(P1-P25) of the FOV respectively. }
\end{figure}

The situation changes when the binning is performed over the higher
layers, seen from the fourth and fifth panels of Fig.4. The
Fraunhofer lines become a little more easily distinguishable now.
More Fraunhofer lines are found without detectable polarizations,
like lines respectively at 519.2nm, 520.2nm, 523.0nm, 523.6nm,
524.2nm, 528.2nm, 528.4nm, 529.7nm, 529.0nm, 530.7nm and 531.5nm.
The style of $Q/I$ profile turns to be simpler with a negative
single valley or a single peak. The polarimetry becomes more noisy
with RMS 6.30$\times 10^{-4}$ greater than 1.72$\times 10^{-4}$ of
the former binning. The amplitudes become larger for these strongly
polarized lines like the magnesium emission triplet and the
Fraunhofer line FeI/CaI527.0nm, while the mixed FeI/CrI/NdII lines
at 529.4nm become too weak to be detected. Another Fraunhofer line
CrI/FeI/FeII527.3nm can be seen from the $I$ profile, its
polarization looks not so prominent but actually larger than that in
the lower layers. On the contrary, the Fraunhofer FeI521.7nm and
FeI/CrI532.4nm lines become more remarkably polarized. It is evident
that in general, the polarization amplitudes of the Fraunhofer lines
become larger than their emission counterparts in the lower layers
for more lines, no critical variation in polarization is found
compared with those pure Fraunhofer lines covering the whole FOV, or
only partial FOV.

Now, let us see the result after the whole FOV binning shown in the
two bottom panels. The $Q/I$ profile looks again closer to that in
the lower layers. But the amplitudes of negative lobes standing
below the continuum polarization level of $Q/I$ profiles of the
quartet are enhanced due to the contribution from the higher layers,
and number of lines with pure positive $Q/I$ amplitudes in the lower
layers is reduced such as FeI/TiII522.7nm line. However, because
that its polarization directions are the same in the two divided
domains, the polarization amplitudes of FeI/CrI527.3nm line becomes
greater than in the lower layers. Finally, it is easily concluded
that different spatial resolutions lead to different polarimetric
results according to the binning schemes. This indicates that the
lines are not homogeneously but variably polarized in space.

\subsection{Spectropolarimetric results of Fraunhofer lines with the coronal line}

Spectropolarimetry of Fraunhofer lines in this section differs from
that of the above-mentioned one only in a way that they share the
same line-of-sight space projected into the sky plane with the green
coronal line, and so do other emission lines. Although it is
difficult to precisely locate these spatial points yielding
Fraunhofer lines due to the projective effect, they can be regarded
definitely as parts of the F-corona when they are formed above the
elongation where the net green coronal line intensity above the
continuum becomes the strongest. The strongest net intensities of
the green coronal line above the continuum locate at P11, P2, P18,
P4 and P5 at columns of the FOV from the left to the right for
SE204-3. For SE205-1, these points with greatest intensities in
these columns of the FOV locates at P11, P2, P3, P4 and P5.
Radiation from above these elongations can be definitely judged in
the corona.

\subsubsection{SE204-3}

The spectral image plotted on top panel of Fig.5 shows SE204-3 case
as one kind distribution of the Fraunhofer lines, where only several
lines have appearance of emission in the lower elongations but
change to absorption one in the larger heights. The FOV of SE204-3
locates at an estimated height of 0.04$R_{sun}$(cf Fig.1), higher
than SE205-1. Although it is found that the Fraunhofer lines keep
roughly their relative intensities as those of the photosphere in
some wavelength interval, but many exceptions are found. For
instance, photospheric lines of the quartet especially MgI516.7nm
and FeI/FeII516.9nm, 518.7nm, 518.8nm, 519.1nm, 520.5nm,
CrI/CrII/FeI527.5nm and FeI/CrI532.4nm become much weaker compared
with their neighboring lines. The transition zone lines at 531.7nm
and the mixed FeII/FeI/NiI519.7nm always take a form of emission as
the green coronal line, but their intensities become weak with
height much more greatly than the latter. At P1, only the green
coronal line, FeII531.7nm and FeII/FeI/NiI519.7nm take a pure
emission appearance, while MgI$b_{4}$ owns a pure absorption profile
and keeps it in the whole FOV. It is interesting that the
MgI$b_{1,2}$ have an emission feature in the line cores and
absorption character in line wings. Moving horizontally along the
first row to the right(cf. Fig.1), the emission feature becomes
stronger and stronger and the absorption weaker and weaker till P5.
On the second row, the emission decays correspondingly. This shows
how the emission appearance emerges and disappears for those lines
with both the partial emission and absorption features.

Since most of these emission lines are confined in the lowest two
rows of the FOV, we perform the first polarimetric demodulation
after the spatial binning over these two rows, i.e., P1-P10. The
second and third panels of Fig.5 provide the corresponding
integrated intensity $I$ and its corresponding fractional linear
polarization $Q/I$ profiles. At first glance, it is noticed that the
polarization level of the continuum becomes now evidently negative.
$Q/I$ profiles of these emission lines in the short wavelength band
are different from those of SE196-2. The $Q/I$ amplitude of
magnesium $b_{2}$ line becomes weaker and the blue wing of its $Q/I$
profile is greatly broadened. The emerged green coronal line gains a
$Q/I$ amplitude of only about 0.07$\%$. Although the FeII531.7nm
lines have intensity larger than the green coronal line, its
polarization is undetectable, unlike the two previous cases.
Meanwhile, its neighboring weak Fraunhofer line FeI531.5nm attains
an impressive polarization as a contrast. Another emission line
FeII/FeI/NiI519.7nm has neither detectable polarization in both the
lower and higher layers.

Now, let us focus on the other Fraunhofer lines. Outstanding
phenomenon occurs that there are at least four appreciable
polarization valleys appearing respectively around 517.7nm, 518.2nm,
525.5nm, 526.0nm and 530.2nm, containing many weak Fraunhofer lines
and including their adjacent continuum. Such kind of polarization
valleys can be also found in SE196-4 in the lower layers around
523.0nm and 531.5nm. It is noticeable that this phenomenon is also
found in the second solar spectrum(Gandofer,2000), say, around
396.0nm, 428.0nm, 435.2nm, 517.2nm, 518.4nm 589.5nm. The
FeI/CrI532.4nm is only one Fraunhofer line whose $Q/I$ amplitude
surpasses 0.1$\%$ wrt the adjacent continuum level. The strong mixed
FeI/CrI lines at 520.4nm and 520.8nm are polarized with considerable
amplitudes, and so do the weak lines at FeII/FeI/CrI526.6nm,
FeI/CaI527.0nm and FeI531.5nm. However, $Q/I$ profiles of the former
two lines gain profiles of approximately antisymmetry. And so do the
magnesium $b_{4}$ absorption line, FeI523.3nm, CrI527.2nm and
FeI/NiI528.2nm. Therefore, such a kind of profile feature is not
owned only by the emission lines. On the other hand, some strong
Fraunhofer lines such as FeI520.2nm, 522.7nm, 525.0nm, 526.3nm, and
FeI/CrI524.7nm have no detectable polarizations. On the whole,
different lines behave distinguishably according to the polarimetric
results, but no critical difference of profile configuration is
found between the emission and Fraunhofer ones. Like the previous
cases, the percent polarization has nothing to do with the line
intensity.

\begin{figure}
\flushright
\includegraphics[width=14.2cm,height=3.2cm]{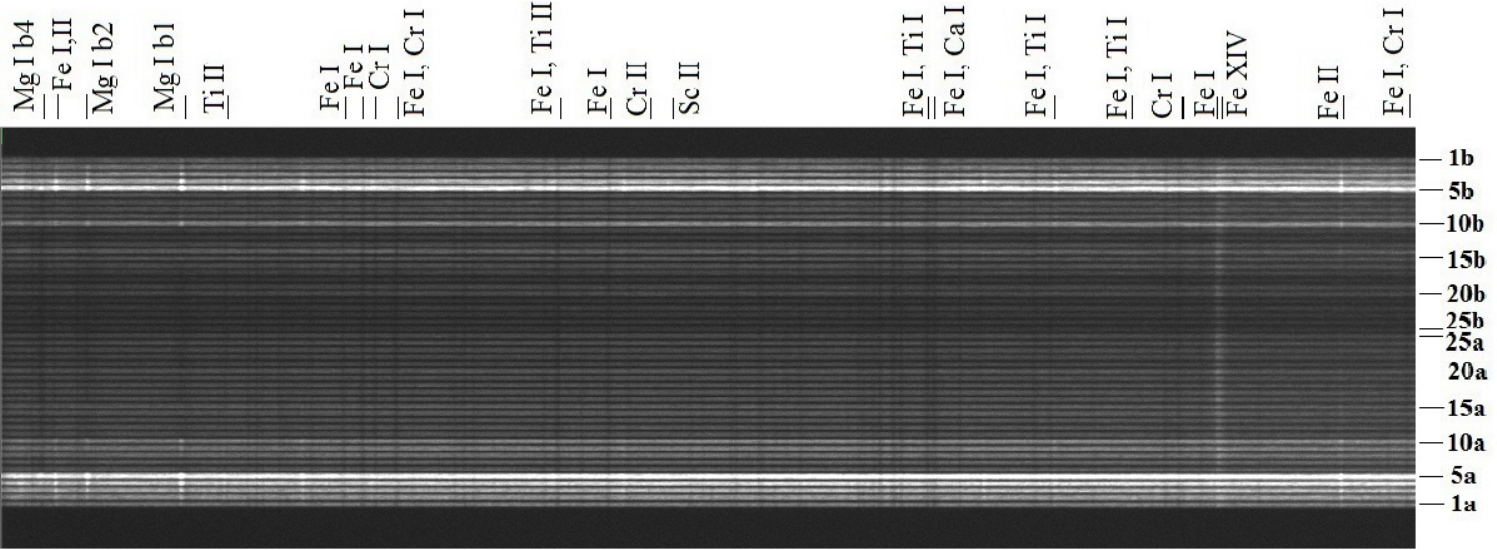}\\
\hspace{3.8cm} \flushleft
\includegraphics[width=16cm,height=5.2cm]{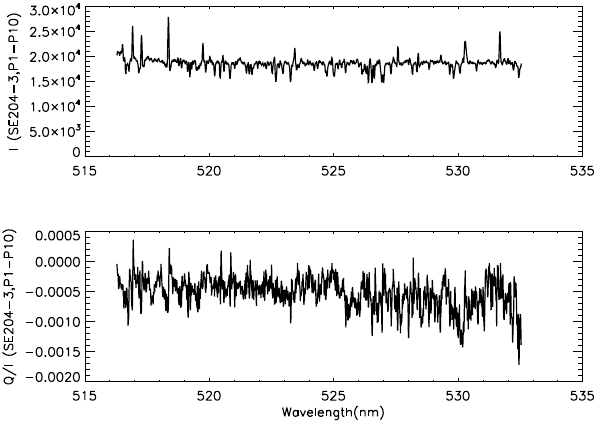}\\
\hspace{3.8cm} \flushleft
\includegraphics[width=16cm,height=5.2cm]{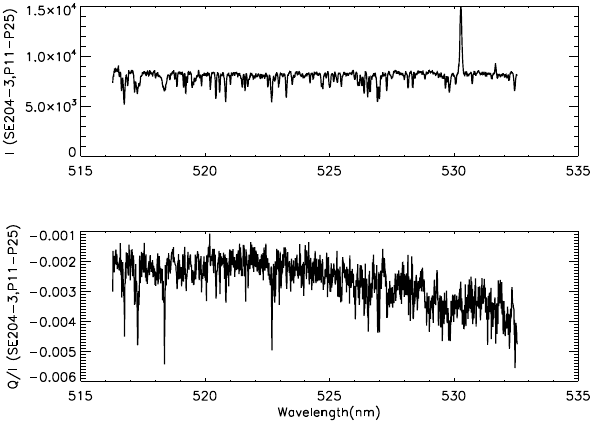}\\
\hspace{3.8cm} \flushleft
\includegraphics[width=16cm,height=5.2cm]{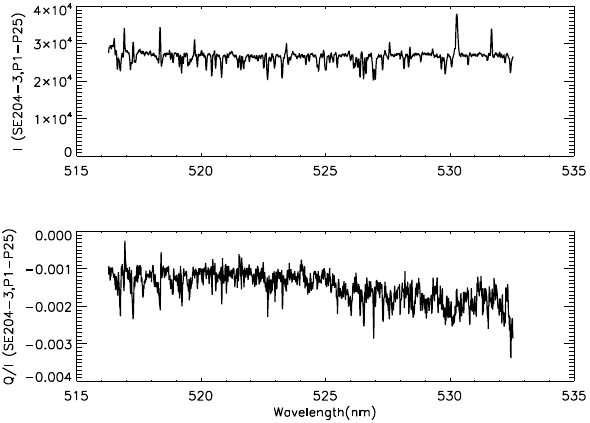}
\caption{\footnotesize SE204-3. Top panel: the raw spectral image.
Lower panels: the intensity $I$ and fractional linear polarization
$Q/I$ profiles obtained from the polarization demodulation after
spatial binning P1-P10, P11-P25 and all the volumes(P1-P25) of the
FOV respectively. The green coronal line is the broadest line at
530.3nm. Note that the polarizations averaged over the upper
layers(P11-P25) are given of F-corona, E-corona and K-corona.}
\end{figure}

The intensity and fractional linear polarization profiles obtained
after binning over the three upper rows are depicted in Fig.5, along
with the spectral image on the top. As before, fewer Fraunhofer
lines are detected to be polarized, but the polarization amplitudes
remarkably increase in these higher layers. The $Q/I$ profiles
depicted in the fifth panel of Fig.5 look similar to those derived
from the former cases in the short wavelength domain. It is clear
that the detectable polarization amplitudes increase in SE204-3
compared to these two above-mentioned cases. A great amplitude of
about 0.36$\%$ is gained by the MgI$b_{1}$ line wrt the continuum
level while in SE194-2 and SE196-4 it is less than 0.25$\%$. The
strong Fraunhofer line FeI/TiII522.7nm obtains prominent
polarization, but such scenario can be found neither in the lower
layers nor in the two previous cases. On the contrary, it is
noteworthy that the Fraunhofer FeI/FeII516.9nm line becomes much
weaker, and so is their polarization. This is very different from
its emission counterparts in the lower layers as an exception of
polarization amplitude enhancement with height. The polarization of
the green coronal line is considerably polarized in the lower region
but seems to be under the detection here. This is due to the
rotation of the polarization direction within these points, as also
seen in Paper I and Paper II, and then canceling out by the spatial
integration. It is noteworthy that the polarimetric results of the
Fraunhofer lines and the continuum in these layers tell us that
polarization amplitudes of F-corona are definitely larger than those
of K-corona.

In summary, the transformation of the emission line to the
Fraunhofer ones seems to change their $Q/I$ profiles and enhance
their amplitudes for some strong lines like the magnesium triplet,
but opposite change happens for other lines such as FeI/FeII516.9nm
which has no detectable polarization in the greater heights, and no
change occurs for other lines like FeII523.4nm and FeI/TiI528.3nm
without polarization signals out of the noise level in both the
domains. For these lines keeping their Fraunhofer profiles in the
whole FOV, some lines attain  polarization detected in the lower
region but lose it in the higher like FeI523.3nm. For strong
Fraunhofer lines like FeI522.7nm, it gains visible polarization in
the higher layers rather than in the lower one. For those Fraunhofer
lines at 520.4nm and 520.8nm, their $Q/I$ amplitudes stand above the
continuum level in the lower region but with greater amplitudes
below the continuum in the higher one. Finally, for these green
coronal line and transition zone line FeII531.7nm, they keep their
emission appearance in the entire FOV. The coronal line is more
polarized in the lower heights and the transition zone line is not
found to be polarized in the lower layers but it seems to have a
little $Q/I$ amplitude in the upper ones. On the whole, the
variation with height is complicated, rather than consistent.

The $Q/I$ profile shapes obtained after binning over all the spatial
points, plotted in the bottom panel, becomes evidently different
from the counterparts of the two previous cases, not only at the
quartet, but also in the long wavelength domain. For instance, $Q/I$
profile of MgI$b_{2}$ line has only a valley here but owns a peak
else in the other two cases. On the whole, the negative amplitudes
below the continuum level dominate in the present case, but these
above the continuum one become fainter. The mixed FeI/CrI/TiI532.4nm
lines have the greatest $Q/I$ amplitude of almost 0.20$\%$ wrt the
continuum level due to absence of cancelation of opposite
polarizations, once again greater than in the above-mentioned cases.
On the contrary, the fractional linear polarization of the green
coronal line FeXIV530.3nm is shown to be almost under detection
after the binning over the whole FOV, again due to cancelations of
$Q/I$ amplitudes with opposite polarization directions in the binned
space. On the other hand, a group of Fraunhofer lines around 527.0nm
attains impressive polarizations. The averaged continuum
polarization amplitudes reach about -0.12$\%$ at the shortest
wavelength to -0.17$\%$ at the longest wavelength.

\subsubsection{SE205-1}

The last case is SE205-1, acquired with an exposure time of five and
half seconds, only nearly eleven seconds before the second contact.
The integrated $I$ and $Q/I$ profiles are depicted just below the
top panel of Fig.6. The Fraunhofer lines exist in a way similar to
SE196-4. For instance, lines at 521.6nm, 521.8nm, 522.5nm and
522.7nm are hardly recognizable in the lower layers. It is not
difficult to find that the relative intensities among the remanent
strong Fraunhofer lines in the upper domain become close to each
other, as evidenced from $I$ profiles plotted in the fourth panel,
very different from the spectra acquired in the quiet sun region
before the eclipse(cf. top panel of Fig.2). It is valuable to note
from the bottom left panel of Fig.1 that just in this case, the
negative $Q$ polarization along the East-West becomes approximately
tangential to the local limb now.

It is evident that the linear polarization profiles obtained after
binning over the two lower rows in the FOV are not so similar to
those of SE196-2 or SE196-4 shown in Figs.3-4, but closer to that of
SE204-3 though its intensity profiles are easily distinguishable
from those of SE204-3 but closer to the previous cases. Now, the
continuum polarization is negative throughout the band and vary from
about 0.07$\%$ in the shortest wavelength to 0.13$\%$ in the
longest. Contrast to many strong emission lines, it is remarkable
that the fractional linear polarizations of most of the Fraunhofer
lines integrated over P1-P10 are not well off the continuum, like
the Fraunhofer lines themselves in the spectral intensity $I$ image.
However, exceptions are present. The large polarization valleys made
by Fraunhofer lines respectively around 518.0nm, 526.9nm and 530.0nm
can be still be seen. At about FeI/CrI532.4nm, the fractional
polarization amplitude wrt the continuum approaches a value of
0.2$\%$. Other Fraunhofer lines like FeI525.0nm line, their percent
linear polarizations can hardly be distinguishable from the
continuum one, while RMS of the polarimetric noise level reads as
3.34$\times 10^{-4}$. In this case, the polarization of the coronal
line is very prominent. Its $Q/I$ amplitude stands up in the large
valleys around it.

\begin{figure}
\flushright
\includegraphics[width=14.2cm,height=3.2cm]{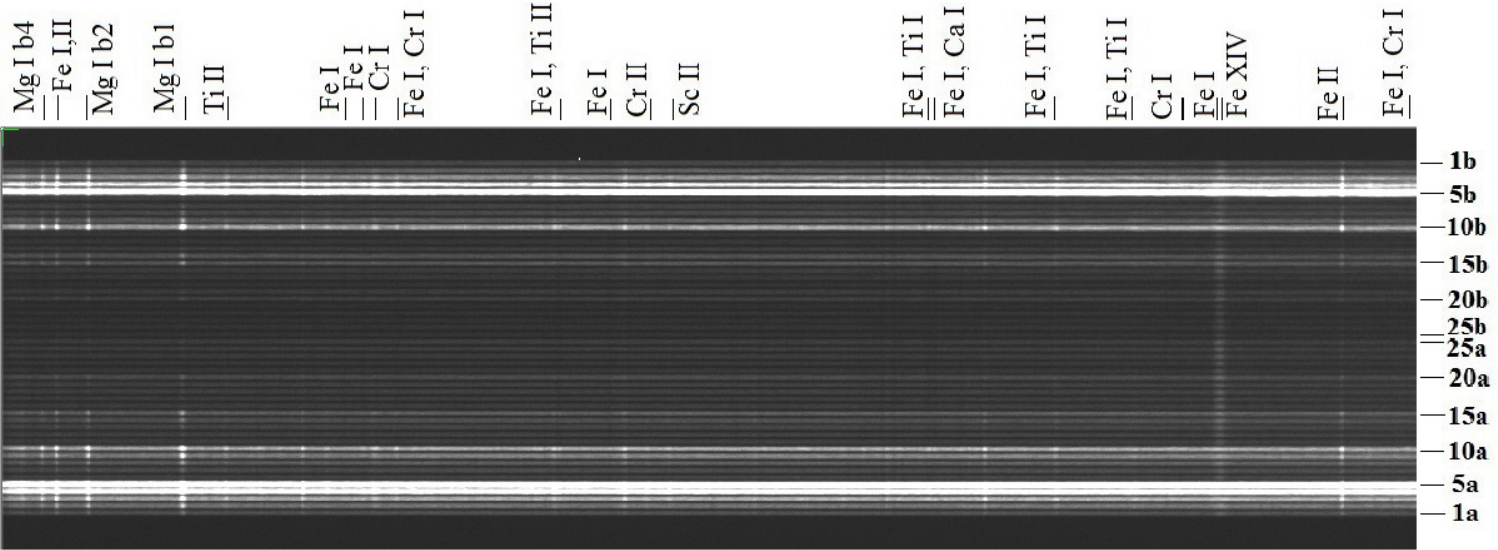}\\
\hspace{3.8cm} \flushleft
\includegraphics[width=16cm,height=5.2cm]{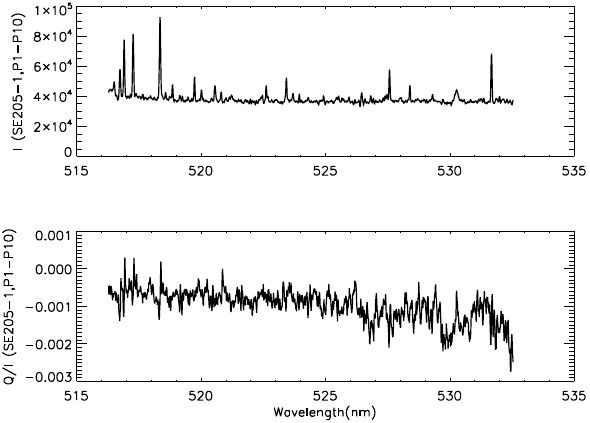}\\
\hspace{3.8cm} \flushleft
\includegraphics[width=16cm,height=5.2cm]{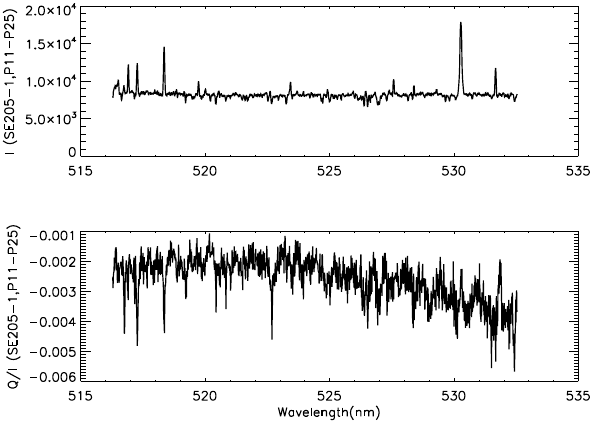}\\
\hspace{3.8cm} \flushleft
\includegraphics[width=16cm,height=5.2cm]{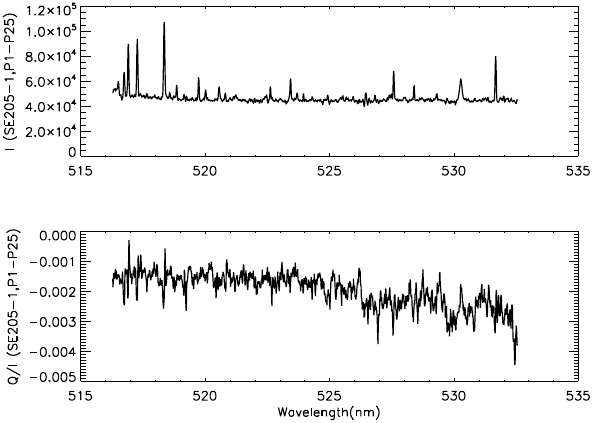}
\caption{\footnotesize SE205-1. Top panel: the raw spectral image.
Lower panels: the intensity $I$ and fractional linear polarization
$Q/I$ profiles obtained in the panels below from the polarization
demodulation after spatial binning over P1-P10, P11-P25 and all the
volumes(P1-P25) of the FOV respectively. Note that the polarizations
are given of F-corona, E-corona and K-corona.}
\end{figure}

Moving to the upper layers formed by P11 to P25 in the FOV, the
situation becomes changed. Both of those Fraunhofer lines themselves
and their detectable polarizations become evidently visible with
considerable fractional linear polarization signals. Polarization of
the Fraunhofer FeI522.7nm line in the upper layers becomes again the
most prominent, like the corresponding situation in upper layers of
SE204-3 but different from the other two cases, with an amplitude
close to 0.30$\%$ as that of MgI$b_{2}$. Those $Q/I$ amplitudes of
the Fraunhofer FeI/CrI520.4nm and 520.8nm are also remarkable,
comparable to the weak emission line MgI$b_{4}$, even stronger than
the green coronal line which now becomes the strongest emission line
and its polarization stands still within the center of the
polarization valley, while it is much weaker in the lower layers
compared with other strong emission lines. The strong emission line
FeI/FeII516.9nm has $Q/I$ polarization less prominent than its
neighbor MgI$b_{4}$ line. Another outstanding polarization is
occupied by emission ScII/CrI 531.8nm line but rather than
FeI/FeII531.7nm. And once again the Fraunhofer FeI/CrI532.4nm line
remains remarkable due to its polarization rather than its poor line
depth. Many Fraunhofer lines have moderate polarization amplitudes,
like these at 519.2nm, 526.9nm, 527.0nm, 530.1nm and 531.5nm. Again,
some Fraunhofer lines, like those at 522.5nm, 523.2nm and 529.7nm,
have no detectable polarization under the present polarimetric
sensitivity of about 6.0$\times 10^{-4}$. Finally, amplitudes of the
continuum polarization become larger from the shortest wavelength to
the largest one by an increase of 0.18$\%$, but they are smaller
than those of most emission lines and most Fraunhofer lines. In
other words, the F-corona and E-corona can be polarized more
strongly than the K-corona.

The $I$ and $Q/I$ profiles of the whole FOV are respectively
depicted in two bottom panels of Fig.6. The intensity profiles are
closest to these of SE196-4 again, except the presence of the
coronal line. But the $Q/I$ profiles look differently from the other
cases. Although more contribution comes from the lower layers to to
the intensity as a result of the combination of the two divided
regions, contribution of negative value of the polarization from the
upper layers cannot be ignored. In fact, all the polarizations are
integrated to be negative or parallel to the local solar limb. For
FeI520.8nm, its totally combined $Q/I$ takes a shape approximately
antisymmetric about the line center, the lobe in the blue wing comes
from the upper layers and the red wing lobe is contributed from the
lower layers. The largest $Q/I$ amplitude of about 0.18$\%$ is
gained again by the Fraunhofer FeI/CrI 532.4nm line again. The
polarization of the coronal line can be easily seen but that of the
transition zone FeII531.7nm lines is almost merged in the noise.

From the above analysis, it becomes evident that some Fraunhofer
lines such as the quartet, CrI/FeI520.8nm, FeI/TiII522.7nm,
FeI523.3nm, and FeI/CrI532.4nm are much more easily polarized than
other lines like FeI523.0nm, FeI/CrI524.7nm and FeI525.0nm.
Furthermore, we have witnessed the great intensity variations of the
same lines in different spatial points. They can be strong
Fraunhofer lines in one spatial point, but hard to be recognizable
in another, such as MgI$b_{4}$, FeI528.2nm, FeI/TiI528.3nm, and
FeI528.8nm. Evidently, this phenomenon cannot be explained via
scattering by dust in the F-corona.

\section{Presence of neutral atoms in the inner corona}

The above-mentioned properties of the Fraunhofer lines in the upper
solar atmosphere revealed from the polarimetry make us search for
the essence of the inner F-corona. It becomes clear that it cannot
be dominantly originated from the dust scattering that should result
in the maintenance of the relative intensities among the Fraunhofer
lines as in the photosphere and the close linear polarimetric
properties for different lines. In fact, even the relative
intensities among these Fraunhofer lines in the inner corona are
observed to be varied greatly from their photospheric counterparts.
For instance, in the spectrum of the quartet in the upper
layers(P11-P25) of the FOV shown in Fig.5, their relative
intensities of the spectral lines are greatly changed from their
photospheric origins, and so do these Fraunhofer lines within
sub-band from 520.0nm to 520.1nm.

Taking SE204-3 for example, the polarizations of the Fraunhofer
lines differs so greatly that they can neither be ascribed to the
dust scattering, since they depend on the wavelength or specific
spectral lines as shown previously. Especially, the co-existence of
$Q/I$ profiles of the approximately antisymmetric shapes with
one-peaked or one-valleyed $Q/I$ profiles of the different
Fraunhofer lines at the same point, further demonstrate that they
cannot be deduced from the line-of-sight integration of the dust
scattering. On the other hand, there are no critical changes of
$Q/I$ amplitudes of Fraunhofer lines, say, FeI532.4nm, from cases
without the coronal line to those with the coronal line.
Furthermore, variation of the $Q/I$ profile of, say, MgI$b_{1}$
line, from the emission(in the lower layers) to the absorption
appearance(in the upper layers) is essentially same as that of
Fraunhofer line keeping their line depression appearance in the
whole FOV, such as FeI520.8nm. All these incline to a conclusion
that these Fraunhofer lines form in the inner corona with small
amplitudes of linear polarization, or the amplitudes will be much
greater(Blackwell,1956) beyond the inner corona since the scattering
becomes more anisotropic with height;

The relative motions can provide further demonstration between two
spatial volumes of these particles yielding the Fraunhofer lines
revealed from the spectra in SE204-3 in the upper layers, derived
from the relative shifts of line center wavelengths along the
line-of-sight due to the Doppler effect.

In order to get the reliable polarimetric result, we have to perform
binning over at least ten spatial points or two rows in the FOV.
However, for purpose of the relative line-of-sight motion
derivation, intensity profile of each spatial point, like these
depicted in Fig.7, can be used from SE204-3, since the standard
deviation of the noise fluctuation reads as 7cts, and the intensity
differences between the line center and its closest wavelength
samples are generally greater than the fluctuation. The relative
motions indicated by Doppler shifts can be reflected in the
wavelength distances between their line centers of these chosen
Fraunhofer lines. The intensity profiles of four sample points P14,
P16, P20 and P22 of SE204-3 in the inner corona are plotted in
Fig.7. The center wavelengths of selected lines are listed in Tab.1
for data analysis.

It is easily seen from Tab.1 that the line center shifts vary for
different lines. For instance, at MgI516.7nm under the present
spectral resolution of 0.012nm, the wavelength displacement of the
line centers between P20 and P22 becomes the greatest as 0.024nm,
the same shift is found at FeI523.3nm, but only 0.012nm for
FeI520.2nm and FeI/CrI520.8nm while no displacement at
FeI/CrI532.4nm but opposite displacement for FeI/TiII522.7nm are
found. On the other hand, the greatest displacement 0.024nm occurs
between P14 and P16 for FeI/CrI532.4nm, but opposite displacement
0.012nm for MgI516.7nm and FeI522.7nm is revealed, while no
displacement occurs for FeI520.2nm, 520.8nm and 523.3nm. Other
similar non-synchronous displacements can be found in this table.
All of these specify that the Fraunhofer lines are not produced
mainly from the dust scattering but individual atoms. These atoms
are not confined in the dust but separately move as individually.

\begin{table}
\caption{Line center positions of sample spectral lines in upper
layers of SE204-3 (wavelengths in the table are in unit of
nanometer)}
\begin{tabular}{lllllll}
\hline \hline
 & MgI516.7nm & FeI520.2nm & FeI/CrI520.8nm & FeI/TiII522.7nm & FeI523.3nm & FeI/CrI532.4nm \\
\hline
P14 & 516.760 & 520.207 & 520.826 & 522.671 & 523.266 & 532.429 \\
\hline
P16  &516.748 & 520.207 & 520.826 & 522.659 & 523.266 & 532.453 \\
\hline
P20 & 516.736 & 520.195 & 520.814 & 522.683 & 523.254 & 532.441 \\
\hline
P22 & 516.760 & 520.207 & 520.826 & 522.671 & 523.278 & 532.441 \\
\hline \\
\end{tabular}
\end{table}

\begin{figure}
\flushleft
\includegraphics[width=17cm,height=6.0cm]{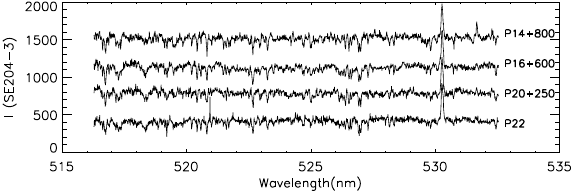}\\
\caption{\footnotesize The intensity $I$ profiles respectively of
P14, P16, P20 and P22 of SE204-3 for illustrating the relative line
center shifts due to Doppler effect. Note that a cosmic event
causing a strong emission at 520.9nm at P22. The intensities of P20
are artificially added with 250cts, and intensities of both P16 and
P14 are added respectively with 600cts and 800cts in order to
separate these four profiles within one panel.}
\end{figure}

\section{Discussion and Conclusion}

In this paper, we present the spectropolarimetry of the Fraunhofer
lines in upper solar atmosphere below elongation of about 0.04 solar
radius including those forming in atom-induced F-corona during a
solar eclipse. Four samples are given with different distribution
style of particles yielding the Fraunhofer lines in the upper
chromosphere, transition zone as well as the inner corona. Here are
the summaries describing the polarimetric results:

1) The polarizations of the Fraunhofer lines and their adjacent
continua can be easily distinguished in the spectropolarimetry. It
becomes evident in cases shown in subsection 2.2 in the upper layers
that the polarization amplitudes of both the E-corona and F-corona
are generally greater than those of the continuum (K-corona)
polarization in the inner corona. And it is evident that those
Fraunhofer lines appeared with the green coronal line have
polarization properties of amplitudes and orientations very close to
those without the coronal lines. This may indicate their common
photospheric origin for these Fraunhofer lines;

2) The polarizations of the Fraunhofer lines described by $Q/I$ vary
with space especially with elongation, in both the polarization
amplitude and orientation. Higher the elongation is, greater the
polarization amplitude is wrt the continuum polarization level as an
overall trend. The distributions of polarizations of all these
Fraunhofer lines are far from the homogeneity or symmetry in the
inner solar corona;

3) The fractional linear polarizations of the Fraunhofer lines in
these layers can reach a few thousandth. A greatest amplitude of
about 0.36$\%$ is detected for the neutral magnesium $b_{1}$
Fraunhofer line in case of SE204-3 after binning over the three
higher rows of the FOV. This is basically consistent with the
prediction by van de Hulst(1950) and observation by Blackwell(1956).
They are often comparable with those of emission lines within the
same FOVs;

4) Different Fraunhofer lines have generally different amplitudes
even in the same volume and their $Q/I$ profiles are structured.
This means that the polarization is wavelength-dependent. There are
some Fraunhofer lines which are more polarization-sensitive, such as
MgI$b_{1,2}$, FeI/CrI532.4nm, than other lines like FeI525.0nm. This
strongly suggests that the polarizations cannot be primarily
ascribed to scattering by dust. Furthermore, the fractional linear
polarization has nothing to do with the line intensity. Otherwise,
spectral resolution becomes crucial especially to show the
approximately antisymmetric $Q/I$ profiles. In such a case,
polarimetry with lower spectral resolution will result in the
underestimation of the polarization amplitudes and lose the detail
of the polarization feature of a specific line;

5) The linear polarization orientation can be changed and the change
process can be accompanied by the $Q/I$ profile configuration
variation. In lower layers of all the cases containing the
Fraunhofer lines, the fractional linear polarization of the
Fraunhofer lines take a form of approximately antisymmetric shapes
and then such an appearance turns to a configuration with amplitudes
either above the continuum polarization level or below it. Another
profile variation signifying the polarization orientation change is
that positive $Q/I$ amplitudes in the lower layers turn to be
negative amplitudes in higher layers;

6) The continuum polarization becomes also greater with height. It
has 0.05$\%$ or smaller percent polarization amplitude very close to
the limb and can reach 0.30$\%$ in the higher layers. And sometimes
it varies with wavelength, but the variation is not considerable.
However, these amplitudes are generally smaller than those of
emission or Fraunhofer lines.

All of these indicate that the mechanisms responsible for the
emanation of linear polarization are very complex, due to
anisotropic scattering among which atomic polarization may be a main
factor leading to spectral line-dependent polarization( Landi
Degl'Innocenti et al., 1983; Stenflo et al., 1997; Trujillo Bueno,
2002). These polarization properties revealed here can help us
understand the physical contents of the outer solar atmosphere
further. For instance, as a byproduct, the polarimetric property
such as strong line-dependence and special $Q/I$ profiles, along
with other above-mentioned evidences, leads us to show the existence
of neutral atoms in the inner corona, where the dust will be
sublimated due to the high energy flux(Boe et al.,2021). Therefore,
the inner F-corona described in this paper is actually induced by
the neutral atoms of metals like iron, magnesium and chromium.

The presence of the neutral atoms like magnesium, iron and chromium
indicates more microstates found in the inner corona than in those
layers below. Thus it means an increase in entropy from the
photosphere to the corona, and it proves further that the corona is
an open but multithermal system. Though contribution from other
sources cannot be excluded like leakage from the magnetically
confined type II spicules and the prominences, SE196-2, SE196-4 and
many similar samples imply that these neutral atoms may originate
dominantly from the photosphere with their profiles modified by the
emission, absorption and Rayleigh scattering along the line-of-sight
accompanying the upflows(Tian,2021). They escape from heating
through chromosphere and transition zone. This extends the space of
the cool matter, such as carbon monoxide molecules discovered in the
chromosphere by Solanki and his cooperators(Solanki,1994). All of
these establish a constraint on the chromospheric and coronal
heating mechanisms.

{\it This work is sponsored by National Science Foundation of China
(NSFC) under the grant numbers 11078005, U1931206}

\end{document}